# Inverse-Designed On-Chip Terahertz Three-Channel Mode and Wavelength Division Demultiplexer

*Faqian Chong, Yulun Wu, Bingtao Gao, Shilong Li, Hongsheng Chen\*, Song Han\**

F. Chong, Y. Wu, S. Li, H. Chen, S. Han
Innovative Institute of Electromagnetic Information and Electronic Integration
College of Information Science & Electronic Engineering
Zhejiang University
Hangzhou 310027, P. R. China
E-mail: hansomchen@zju.edu.cn; song.han@zju.edu.cn

F. Chong, Y. Wu, B. Gao, H. Chen, S. Han
State Key Laboratory of Modern Optical Instrumentation
ZJU-Hangzhou Global Scientific and Technological Innovation Center
Zhejiang University
Hangzhou 310200, P. R. China
E-mail: hansomchen@zju.edu.cn; song.han@zju.edu.cn

H. Chen
International Joint Innovation Center
The Electromagnetics Academy at Zhejiang University
Zhejiang University
Haining 314400, P. R. China
E-mail: hansomchen@zju.edu.cn

H. Chen
Key Lab. of Advanced Micro/Nano Electronic Devices & Smart Systems of Zhejiang
Jinhua Institute of Zhejiang University
Zhejiang University,
Jinhua 321099, P. R. China
E-mail: hansomchen@zju.edu.cn

# Abstract

High-performance multimode/multiwavelength (de)multiplexer is one of the most pivotal photonic devices for advanced on-chip interconnect systems. Traditional on-chip photonic (de)multiplexing requires large device footprint for maintaining high efficiency, large operation bandwidth, and small insertion losses. Here a hybrid inverse design method is therefore proposed to combine genetic algorithms (GA) and topology optimization for developing an ultracompact (lateral size< 2λ) terahertz (THz) mode-/wavelength-division demultiplexer (MDM-WDM). The method leverages the global search capability of GA in continuous parameter spaces and the local topology optimization strategy driven by the adjoint method, effectively improving design convergence efficiency and global performance robustness. Experimental results demonstrate that the device simultaneously achieves stable three-channel MDM and WDM with insertion loss (IL) of less than 3 dB and inter-channel crosstalk (CT) can be -22 dB. The output achieves spatial separation of the orthogonal $TE_{10}$, $TE_{20}$, and $TE_{30}$ modes that correspond to three central wavelengths $\lambda_1\sim690$ μm, $\lambda_2\sim700$ μm, and $\lambda_3\sim710$ μm, respectively, verifying the device's precise control over target modes and wavelengths. This work provides an efficient optimization approach for developing broadband, multichannel, and highly integrated THz multiplexing devices, offering new pathways for constructing next-generation integrated photonic interconnects and signal processing systems.

# 1. Introduction

With the rapid development of high-performance computing applications such as artificial intelligence (AI) [1-3], hyperscale data centers are growing exponentially [4-6]. This trend poses significant challenges to traditional copper-based electrical interconnect technologies, which struggle to meet the future demands of bandwidth density, low transmission, and low interconnect energy consumption. Optical interconnect technology, due to its superior bandwidth scalability, low energy consumption, and low latency, is considered a core solution to address these challenges [7, 8]. Optical interconnects based on photonic integrated circuits (PICs) can leverage multiplexing technologies to significantly increase data transmission bandwidth density, playing a critical role in chip-to-chip interconnects [9]. Multiplexing technologies, such as wavelength-division multiplexing (WDM) and mode-division multiplexing (MDM), are primarily used to substantially enhance data transmission capacity in data centers and long-haul coherent communications [10, 11]. In optical interconnect architectures, the deep integration of MDM and WDM is considered as a key enabling technology for further expanding link capacity and photonic information dimensions [12-14]. Achieving this requires effective control over modes within waveguides, demanding not only high selectivity for different modes but also minimal crosstalk (CT) between them to ensure signal integrity and system stability. Realizing joint MDM-WDM multiplexing in the terahertz (THz) frequency band (0.1–10 THz) holds significant research value. The THz band provides potential spectrum resources for future ultra-high-speed inter-chip and on-chip interconnects, with wavelengths ranging from hundreds of micrometers to millimeters, offering ample propagation space for high-density mode multiplexing [15-19].

Combining the advantages of the THz band with MDM and WDM technologies can significantly enhance data transmission capabilities. Therefore, developing MDM and WDM devices in THz band that combines wide operating bandwidth, strong mode controllability, low CT, and compact size has become a critical technical bottleneck in photonic chip research. Current mainstream high-order mode demultiplexers typically rely on phase-matching principles to achieve mode conversion [20, 21]. While these designs offer some physical intuitiveness and have successfully achieved multiplexing of multiple orthogonal modes, they generally depend on cascaded structures of multiple mode converters, leading to exponential device size growth with increasing mode orders. Traditional analytical theories and intuition-driven methods struggle to simultaneously satisfy complex constraints such as multi-objective optimization, high-dimensional design, and manufacturing feasibility, necessitating new paradigms.

Recently, inverse design methods, such as target-driven and algorithm-guided structural optimization strategies, have been widely applied in integrated photonics [22-28]. These methods utilize numerical optimization algorithms to efficiently search the parameter space within constraint boundaries, automatically identifying performance-limiting structures, particularly for high-degree-of-freedom problems. Therefore, the inverse design offers greater flexibility, enabling precise control over electromagnetic (EM) wave manipulation and serving as a key approach to breaking through traditional design limitations. Related studies have demonstrated that structures based on inverse design exhibit excellent performance and scalability in optical communications [29-32] optical computing [33, 34], optical sensing [35], and imaging systems [36].

In this work, we propose an all-silicon MDM-WDM device based on a combination of genetic algorithms (GA) and gradient-based topology optimization. This approach achieves a complete optimization workflow from initial structure generation to high-precision performance refinement. GA is used to construct diverse, manufacturable initial structural populations in a continuous parameter space, identifying near-optimal structures through fitness-driven evolution. Subsequently, topology optimization performs continuous refinement based on the adjoint method, further enhancing device performance and enforcing physical constraints. This strategy effectively mitigates the sensitivity of topology optimization to initial conditions, improving convergence rates and global performance. The designed device incorporates auxiliary photonic crystal waveguide (PCW)-based packaged structures, enabling low-loss coupling and stable interfacing with standard THz waveguide systems, ensuring experimental feasibility and mechanical compatibility. The photonic crystal (PhC) package adopts a modular design, enhancing fabrication compatibility and offering excellent potential for system integration. The proposed design framework is suitable for structural optimization and integrated implementation of multiport, multimode, and multiband on-chip photonic devices, particularly as a core demultiplexing module in THz interconnect systems. It provides a scalable design paradigm for future broadband, high-density, and manufacturable integrated photonic platforms.

## 2. Methods and Results
2.1 Design Concept of the Three-channel MDM and WDM

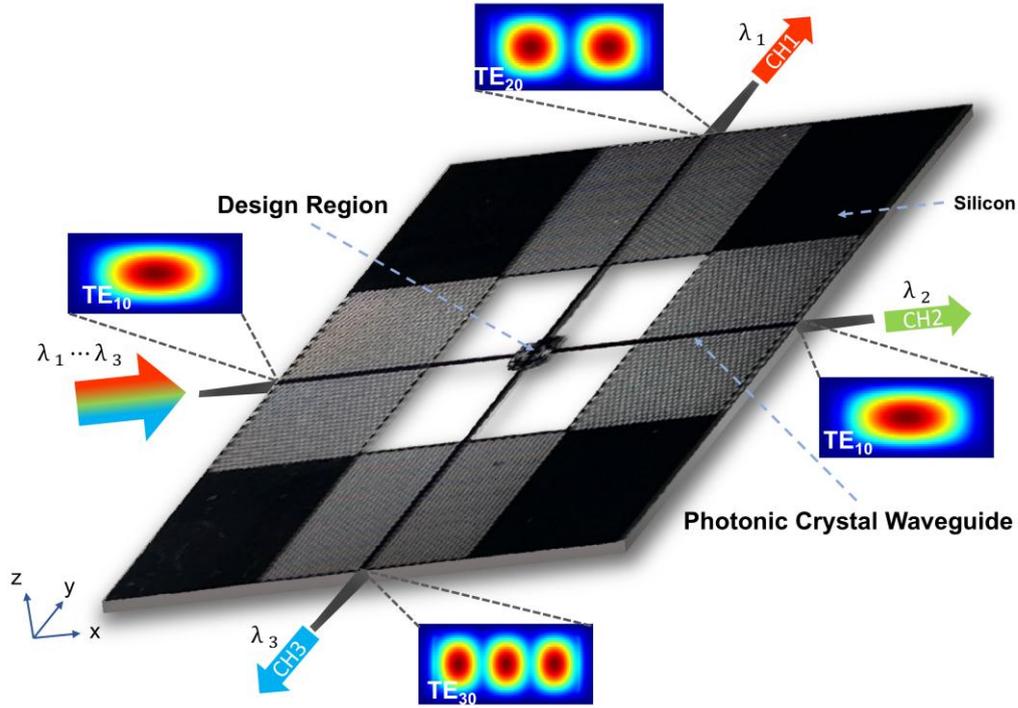

**Figure 1. The THz Photonic MDM-WDM device consists of an inverse design region and a supporting framework based on photonic crystal waveguide (PCW) theory.** At the input, a mixture of the $TE_{10}$ mode and EM waves within the 680-720 μm wavelength range is coupled into the device. At the output, the first port outputs the $TE_{20}$ mode with a center wavelength of 690 μm, the second port outputs the $TE_{10}$ mode with a center wavelength of 700 μm, and the third port outputs the $TE_{30}$ mode with a center wavelength of 710 μm.

The MDM-WDM photonic device consists of a functionally designed region and a supporting framework, which is based on the PCW theory and provides mechanical support to the device. **Figure 1** illustrates the schematic of a three-channel MDM-WDM device based on the integration of inverse design and PCW. At the input of the device, EM waves in the 680 to 720 μm wavelength range are coupled with the $TE_{10}$ mode. At the output, the first port outputs the $TE_{20}$ mode at a center wavelength of 690 μm, the second port outputs the $TE_{10}$ mode at 700 μm, and the third port outputs the $TE_{30}$ mode at 710 μm, respectively.

The functional design region is patterned on a suspended high-resistivity silicon wafer. To achieve efficient coupling, a tapered waveguide is employed at each port. The PCW region between the tapered waveguide and the inverse design region plays a crucial role, specifically in the following aspects: (1) It provides mechanical support to stabilize the tapered input waveguide and the inverse-designed device; (2) The photonic bandgap in this region suppresses lateral mode leakage within the target frequency range, effectively confining the THz wave to the central guiding region through interference effects; (3) A silicon region without PhC patterning is introduced laterally outside the PCW to facilitate device fabrication and experimental handling. This not only improves the fabrication adaptability of the device but also facilitates potential large-scale device integration. By combining inverse design with PCW technology, this approach optimizes both THz wave transmission efficiency and mode control, enhancing the system's wavelength and mode multiplexing capabilities.

## 2.2 Inverse Design Framework

To enhance the efficiency of structural optimization in complex photonic devices, we propose a hybrid design method, which combines GA with gradient-based topology optimization, making it suitable for large parameter spaces. This approach is particularly suited for high-performance inverse design of multi-channel MDM-WDM devices, enhancing manufacturability and providing structural interpretability, which are critical for practical implementation and optimization. The overall algorithm workflow consists of two main steps: continuous optimization with GA and the adjoint method that is empowered by the gradient-based topology optimization, as shown in **Figure 2**.

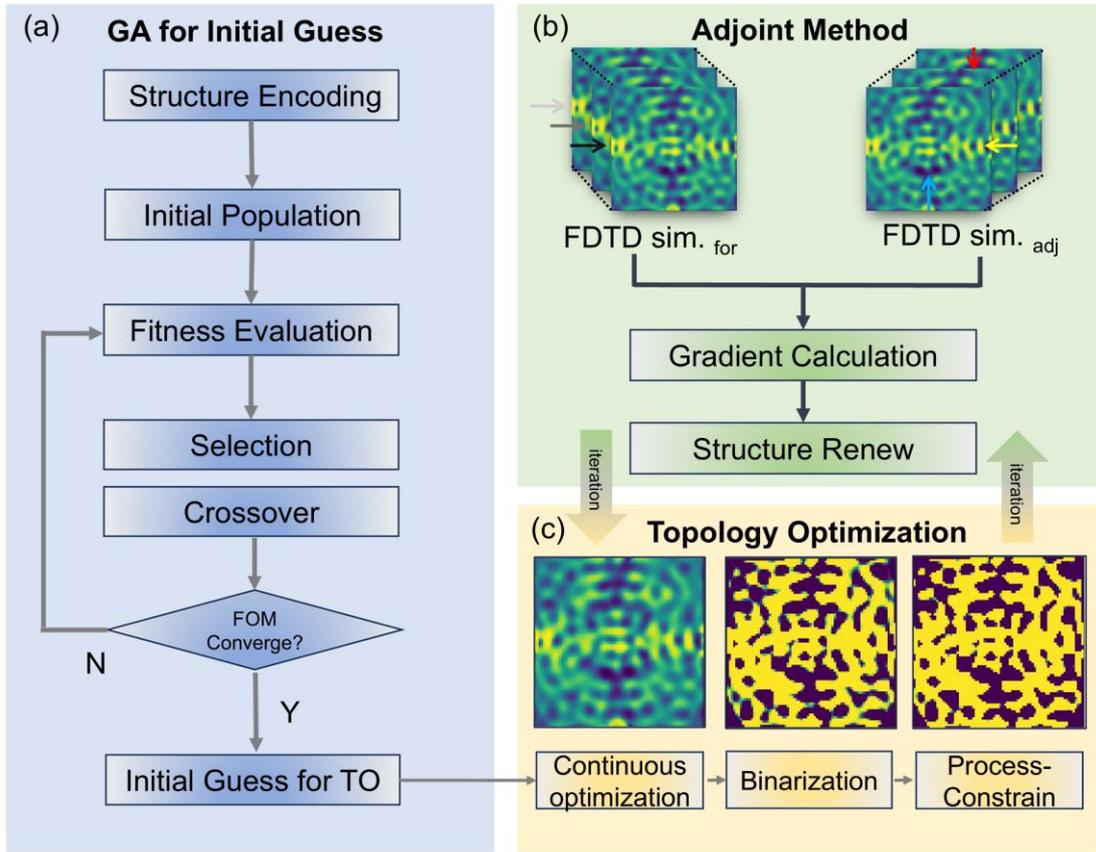

**Figure 2. The inverse design framework.** (a) Continuous optimization using GA, (b) the adjoint method for computing gradients and iteratively updating the structure during the topology optimization process, and (c) gradient-based topology optimization.

First, the device's design region is set to 0.9 × 0.9 mm² and discretized into a 300 × 300 grid for isotropic permittivity. Each grid cell is parameterized as a continuous fill factor variable $x_i \in [0,1]$, where $x_i = 1$ represents a cell fully filled with silicon, or $x_i = 0$ represents air, forming the design vector $x = [x_1, x_2 \cdots x_n]$. In the initial structural exploration phase, a continuous permittivity parameter space is employed, facilitating smooth and differentiable optimization, which enhances the convergence and performance of the design. This continuous parameter space ensures efficient exploration of the design space and avoids abrupt transitions between material states, leading to a more effective optimization process.

Rather than proceeding directly to topology optimization (TO), we recognize the importance of an initial guess in structural optimization. Like the role of pretrained structures in neural networks, the quality of the initial guess significantly affects the convergence speed and quality of the final solution [37, 38]. If the initial structure is overly random or physically unreasonable, the optimization process is likely to converge to local optima, leading to premature convergence. To address this issue, we introduce GA [39] as a structural prior generation mechanism before the topology optimization step, enabling the optimization process to start with a more reasonable initial structure. Through intelligent search, GA generates continuous structures that exhibit strong initial performance, significantly reducing search time during optimization and improving the global search capabilities. The first step, continuous optimization with the GA is shown in Figure 2a, the GA performs continuous optimization. Each population is an array representing the pixel distribution of the device, with the array length equal to the total number of pixels, and each element corresponding to a pixel. At the start of the GA phase, an initial population is defined as $P^{(0)} = [x^{(1)}, \cdots, x^{(M)}]$. The initial population consists of a mix of heuristic structures, such as fully random structures, mirror-symmetric structures, laterally or longitudinally graded structures, and central block structures. This diversity ensures coverage of representative regions of the design space. Each of these heuristic structures is chosen to explore different potential design configurations, enabling a more comprehensive search of the design space.

To accurately evaluate the modal coupling efficiency at each output port, this work adopts a mode overlap power-normalized merit function based on a single channel as the transmittance:

$$T_j = \mu \frac{\left| \int_{S_j} E(r) \times H^*_{m,j}(r) \cdot d\mathbf{S} + \int_{S_j} E_{m,j}(r) \times H^*(r) \cdot d\mathbf{S} \right|^2}{Re\left( \int_{S_j} E_{m,j}(r) \times H^*_{m,j}(r) \cdot d\mathbf{S} \right)} \quad (1)$$

where $E$ and $H$ are the simulated electric and magnetic field distributions, respectively, $E_{m,j}$ and $H_{m,j}$ are the normalized mode field distributions of the target mode at the j$^{th}$ output port. The numerator represents the complex conjugate overlap integral between the actual field and the target mode, which quantifies the degree of mode matching, while the denominator is a normalization factor for the power of the target mode. Since the device has three independent output ports, each demultiplexing different wavelength bands or modes, the merit function for the three ports is combined into a summed form. This ensures that each port's contribution to the overall performance is adequately accounted for in the optimization. To achieve performance control over a broad spectral range, a generalized p-norm merit function is used to perform integrated weighted optimization across the entire band. The optimization figure of merit (FOM) is:

$$FOM = \sum_{j=1}^{3} \left[ \left( \frac{1}{\lambda_2 - \lambda_1} \int_{\lambda_1}^{\lambda_2} (T_j)^p \, d\lambda \right)^{1/p} \right] \quad (2)$$

where $\lambda_1$ and $\lambda_2$ denote the lower and upper bounds of the target optimization wavelength range at the corresponding output port. The FOM is used as fitness function that guides the search for optimal designs in the GA phase of the design process. By integrating this FOM into the GA optimization loop, the device is iteratively refined to achieve high transmission efficiency and minimal CT across the targeted wavelength bands and modes. During individual evolution, tournament selection is used to select parents, and offspring are generated using weighted linear crossover. A Gaussian perturbation mutation mechanism is introduced to promote structural diversity, enhancing the algorithm's ability to explore diverse design solutions. The crossover operation between parents $x_A$ and $x_B$ is expressed as:

$$x_{\text{child}} = \alpha x_A + (1-\alpha)x_B \tag{3}$$

where $\alpha$ is a weighting factor. This helps to combine features of both parents, creating offspring that inherit characteristics from both parents. Additionally, the mutation mechanism introduces small random perturbations to the design factor $x_i$ as shown below:

$$x_{i+1} = x_i + \epsilon, \epsilon \sim N[0, \sigma^2] \tag{4}$$

where $\epsilon$ is a random variable drawn from a Gaussian distribution with mean 0 and variance $\sigma^2$, and $i$ denotes the iteration index in the GA optimization process. The notation $N[0, \sigma^2]$ indicates that $\epsilon$ follows a Gaussian distribution with a mean of 0 and variance $\sigma^2$. This mutation mechanism introduces diversity by perturbing the design parameters in a controlled manner. The optimization process terminates when if no obvious improvement is observed after 15 generations. Otherwise, fitness evaluation, crossover, and mutation continue. During optimization, the population size, crossover probability, and mutation probability are set to 50, 0.9, and 0.1, respectively. After iteration, the structure with the highest fitness is selected as the candidate structural prior. This structure possesses reasonable physical properties and initial performance, providing a more effective starting point for subsequent gradient-based topology optimization.

After obtaining a high-quality initial structural guess from the GA, this work proceeds to the second step: the topology optimization phase. The goal is to refine the geometric structure for improved optical performance, enhance modal control to reduce CT, and improve manufacturability to facilitate practical fabrication. The entire topology optimization process is divided into three stages: continuous optimization, binarization, and process-constrained optimization, as shown in Figure 2c. In the topology optimization phase, the FOM remains consistent with the fitness function used in the GA phase. As the initial stage of topology optimization, the goal of the continuous optimization stage is to optimize the permittivity distribution of each pixel in a continuous space. The intermediate structures generated during continuous optimization contain values between 0 and 1, which are not suitable for manufacturing. After convergence, a binarization operation discretizes the continuous dielectric constants into a binary structure. To ensure the manufacturability of the generated structure in nanolithography and etching processes, a process-constrained stage is implemented, where a circular spatial blur filter is applied at fixed iteration intervals to suppress the generation of small feature sizes.

Efficiently calculating structural gradients in a high-dimensional parameter space is critical for topology optimization. In this process, we adopt the adjoint gradient method to avoid computing partial derivatives for each pixel individually, significantly reducing simulation overhead. Topology optimization relies on adjoint gradients, with gradients computed through forward and adjoint field simulations in each iteration (see Supplementary Note S1 for more details). The structure variables are updated using a gradient descent strategy to approach the target performance [40], as shown in Figure 2b. The iteration process is described by the following equation:

$$\varepsilon_r^{(k+1)}(x) = \varepsilon_r^{(k)}(x) - \eta \cdot \frac{\partial FOM}{\partial \varepsilon_r^{(k)}} \tag{5}$$

Where $\varepsilon_r^{(k)}$ represents the permittivity of the design vector $\boldsymbol{x}$ at the k-th topology optimization iteration. After the topology optimization phase, the device ultimately forms a binary structure composed of only silicon and air. Through continuous permittivity optimization, binarization, and process-constrained optimization, the design gradually meets the target performance, ensuring high efficiency, low loss, and effective modal control.

## 2.3 Device Fabrication and Experimental Method

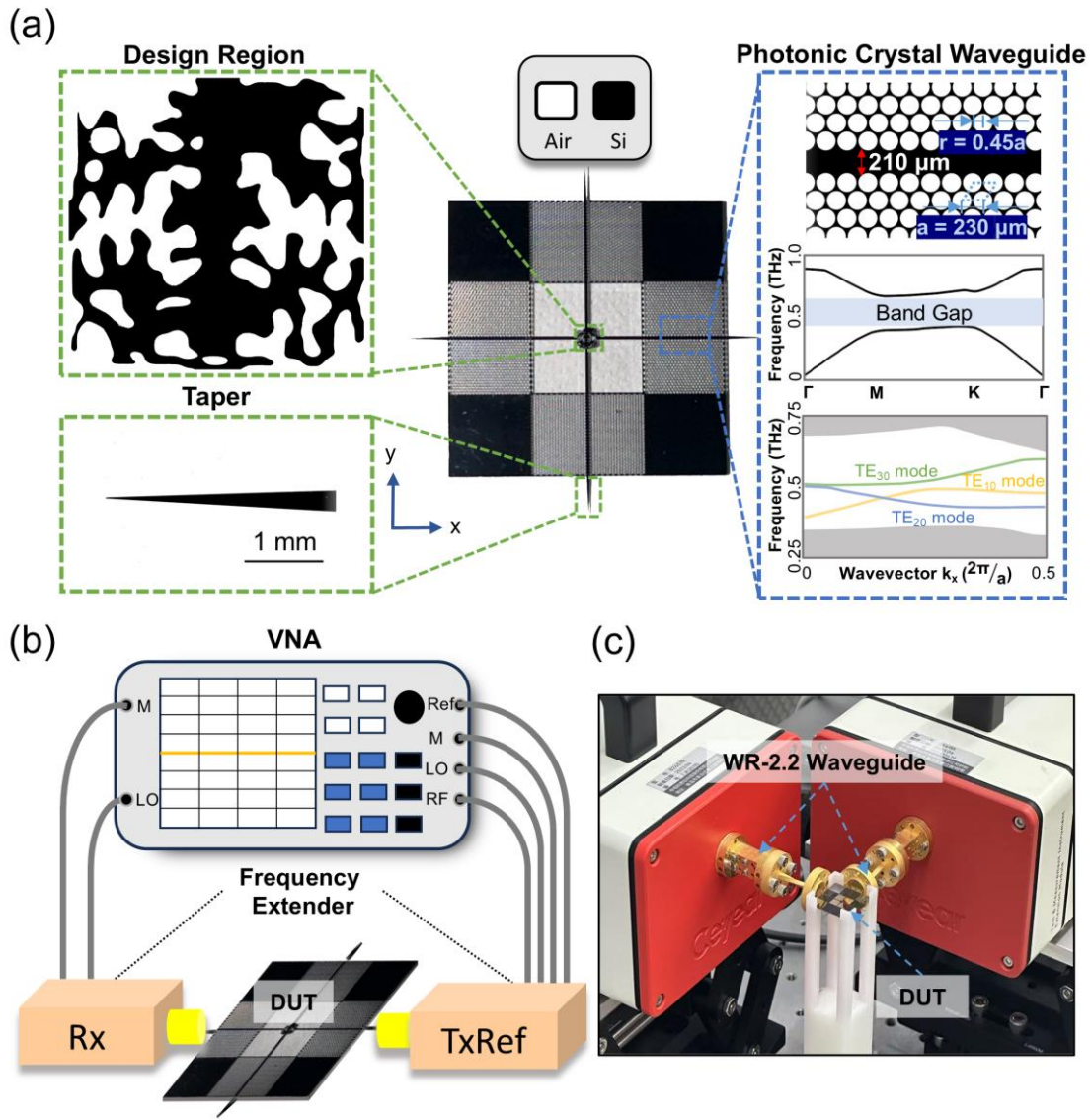

**Figure 3. Fabricated device and experimental characterization setup.** (a) Microscope image of the MDM-WDM device. The left column shows the pattern of the inverse-designed region and the tapered coupler for coupling in-plane TE modes, while the right column presents the pattern of the photonic crystal waveguide (PCW) with a line defect, a schematic of the unit cell bandgap, and the three TE modes supported by the PCW. (b) Schematic of the experimental setup, illustrating the THz continuous-wave measurement system comprising a vector network analyzer (VNA) and a 330–500 GHz frequency extension module. (c) Experimental configuration for near-field characterization of the device. The extension module is connected to the sample via WR-2.2 waveguides, enabling both excitation and collection of the signal through the same waveguide interface.

To achieve efficient modal conversion and stable coupling from standard waveguides to the freeform structure, the device adopts a modular design, consisting of three functional regions. The central topology optimization region, which is the core of the design, is responsible for implementing multi-band modal demultiplexing function. This device is connected to the testing equipment via tapered waveguides, which effectively perform field transformation and match standard waveguide modes to the internal modes of the topology-optimized structure, and reduce reflection losses. The topology optimization region interfaces with the tapered waveguides through a PCW [41], providing excellent bandgap control and propagation stability and ensuring high modal overlap efficiency with the WR-2.2 standard waveguide. The device edges include reserved clamping areas to facilitate mechanically stable fixation using waveguide probes on the testing platform, and minimizes measurement errors. This structure not only simplifies sample installation but also enhances repeatability under various experimental conditions. The modular partitioned design allows the device to connect directly to standard THz testing platforms without requiring additional conversion structures, and greatly improves the compatibility and measurement efficiency of the testing system.

The inverse-designed THz photonic devices were fabricated using standard photomask lithography to define the photonic pattern on a thick photoresist layer (AZ4620, ~7 μm). The pattern was directly transferred onto the silicon substrate using deep reactive ion etching (DRIE) [42]. The devices were patterned on a high-resistivity (>10,000 Ω·cm) single-crystal silicon wafer with a thickness of 200 μm (see Supplementary Note S2 for more details). **Figure 3a** presents the optical microscope image of a representative fabricated device, which included a tapered waveguide coupler, the inverse-designed region (left column in Figure 3a) and a line-defect waveguide with a width of 210 μm embedded in a PhC slab with a TE-like photonic bandgap (right column in Figure 3a). The inverse-designed device exhibits a freeform geometry, with the smallest feature size exceeding 15 μm. This ensures compatibility with DRIE through a 200 μm-thick silicon slab, and maintained an etching aspect ratio larger than 10:1.

To experimentally characterize the performance of the inverse-designed device, a THz vector network analyzer extension (THz VNAX) system is employed, Figure 3b illustrates a schematic of the measurement setup. The testing platform consists of a vector network analyzer (VNA) and its corresponding frequency extension modules, which cover a frequency range from 330 GHz to 500 GHz. The THz generation and detection setup incorporate two electronic frequency extension modules—one configured as a Transmit-Reference (TxRef) module and the other as a Receive-only (Rx) module. These modules interface with a four-port vector network analyzer (VNA), which simultaneously serves as the local oscillator (LO), radio frequency (RF) source, and spectrum analyzer for the measurements. The system uses standard WR-2.2 waveguides for signal transmission. The VNA generates swept-frequency signals, which are sent to the device under test (DUT), while the response signals are collected by the same type of waveguide and returned to the VNA for amplitude and phase measurements of the multi-port S-parameters. During testing, the device sample is securely fixed on a 3D-printed sample holder, ensuring stable and repeatable contact between the waveguide and the tapered waveguide, while Figure 3c provides a photograph of the actual system. This setup minimizes measurement errors and ensures consistent experimental conditions.

## 2.4 Experimental Results and Analysis

**Figure 4a** presents both the simulated (dashed lines) and experimentally measured (solid lines) transmission spectra of the proposed MDM-WDM device across the three target channels. The IL for each channel is quantitatively assessed by computing the modal conversion efficiency using power-normalized mode overlap integrals between the incident input mode and the corresponding target output mode. Inter-channel CT is determined as the minimum amplitude difference, at each channel's center wavelength, between the target channel and the strongest undesired channel response. Experimental results indicate that the device exhibits a relatively flat spectral response over the designed bandwidth, with an IL of approximately 3 dB for all channels and minimum inter-channel CT reaching -22 dB, validating the efficacy of the optimization strategy.

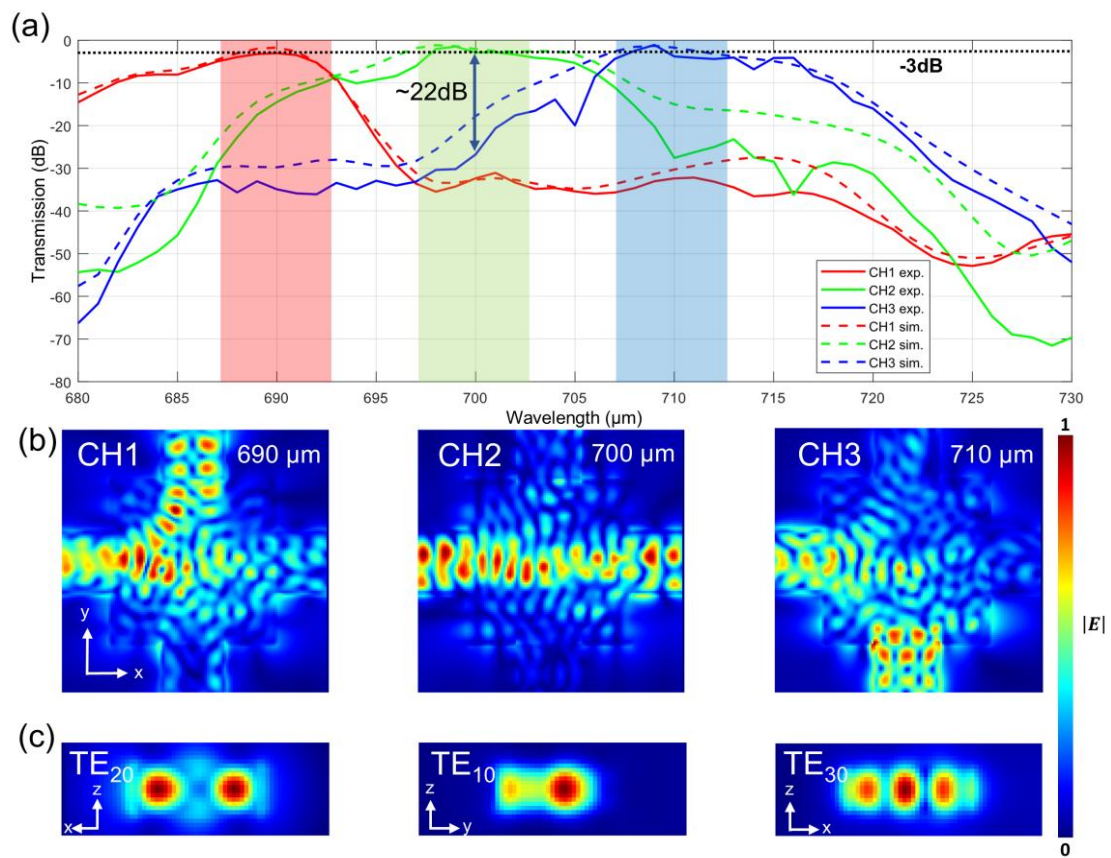

**Figure 4. Simulated and experimental results.** (a) Simulated (dashed lines) and experimental (solid lines) transmission spectra at wavelengths of 690 μm (red), 700 μm (green), and 710 μm (blue). (b) Normalized simulated electric field distributions (|E|) at 690 μm, 700 μm, and 710 μm, respectively. (c) Simulated mode profiles at the output multimode waveguide for each channel at its corresponding center wavelength.

Figure 4b illustrates the EM field manipulation achieved by the topology-optimized structure within the design region. The THz signal flow and field distributions at representative wavelengths are visualized to demonstrate how the device routes energy toward different output ports. Normalized electric field simulations for input wavelengths of 690 μm, 700 μm, and 710 μm clearly show that each signal is efficiently directed to a distinct output port, confirming the simultaneous wavelength and mode separation functionality. While high modal confinement is achieved, minor energy leakage into adjacent channels is observed, contributing to residual CT effects. Residual CT in the topology-optimized device can be attributed to several physical factors. First, limitations in mode orthogonality and photonic bandgap control can lead to partial overlap of modal field distributions. Structural discretization, finite simulation resolution, and restricted design degrees of freedom may prevent perfect orthogonality between the target modes, while the PCW bandgap cannot fully suppress all non-target modes across the entire operating bandwidth, especially near the band edges. Second, incomplete coupling and mode matching between the standard WR-2.2 waveguide and the topology-optimized structure can introduce unwanted signal leakage. Even with tapered waveguides to reduce reflections, residual reflected energy may re-couple into other channels, and mismatch between the incident and target mode profiles can further increase CT. Third, in a compact device layout, free-space radiation and edge scattering can couple energy into neighboring channels. Such non-guided leakage can cause spatial overlap between channels, thereby degrading isolation performance. Figure 4c shows the simulated output modal profiles in the multimode waveguide at the center wavelength of each respective channel. These modal field patterns further verify that the target $TE_{10}$, $TE_{20}$, and $TE_{30}$ modes are successfully and selectively excited at the corresponding ports, demonstrating the device's high fidelity in mode sorting and spectral discrimination.

## 3. Conclusion

This study presents and experimentally validates a broadband, high-performance MDM-WDM design framework for the THz frequency range. The proposed approach integrates GA with gradient-based topology optimization, combining the global search capability of GA with the local refinement capability of gradient-based methods, while satisfying multiple physical constraints. The PCW packaging structure enables low-loss coupling between input/output ports and external testing platforms while preserving the EM performance of the inverse-designed region. This approach ensures high experimental repeatability and facilitates system-level integration. The device is fabricated using standard silicon-based micro-nano fabrication processes, ensuring excellent manufacturability and compatibility with existing technologies.

Simulated and experimental results show that the designed device exhibits a stable spectral response across 690–710 μm and effectively separates the $TE_{10}$, $TE_{20}$, and $TE_{30}$ orthogonal modes at the output, achieving precise control over modal overlap. Simulated power flow density and field distribution profiles further confirm that the structure facilitates precise THz field control and beam shaping, significantly suppressing non-target mode leakage and adjacent channel interference. In summary, the proposed inverse design framework and encapsulation strategy provide a versatile, scalable solution for realizing on-chip MDM-WDM devices in the THz range. The device's compact footprint (<2λ), stable performance, and scalability make it a promising candidate for high-throughput, low-crosstalk on-chip interconnect modules. This design paradigm offers a robust pathway toward high-density, high-performance THz communication and data processing systems, positioning it as a promising candidate technology for future communication infrastructure.

## Supporting Information

Supporting Information is available from the author.

## Acknowledgments

S. H. acknowledges the National Key R&D Program of China (2024YFB2808200), the National Natural Science Foundation of China (62475230), the Excellent Young Scientists Fund Program (Overseas) of China, and the Fundamental Research Funds for the Central Universities.

## Conflict of Interest

The authors declare no conflict of interest.

## Author Contributions

**Faqian Chong**: developed the algorithm, conducted simulations and experiments, and mainly wrote the manuscript. **Yulun Wu**: assisted in experiments and manuscript writing. **Bingtao Gao**: helped with the experimental work. **Shilong Li**: guided the theoretical work and simulations. **Hongsheng Chen**: co-supervised the entire project. **Song Han**: revised the manuscript and supervised the entire project. All authors participated in analyzing data and approved the final version.

## Data Availability Statement

The data that support the findings of this study are available from the corresponding authors upon reasonable request.

## References

1. J. Pei et al., "Towards artificial general intelligence with hybrid tianjic chip architecture," *Nature* **572**(7767), 106-111 (2019). https://doi.org/10.1038/s41586-019-1424-8.
2. L. Jin et al., "Generalized phase retrieval model based on physics-inspired network for holographic metasurface (invited paper)," *Progress In Electromagnetics Research* **178**, 103-110 (2023).
3. Y. Feng et al., "Optical neural networks for holographic image recognition (invited paper)," *Progress In Electromagnetics Research* **176**, 25-33 (2023).
4. Q. Cheng et al., "Recent advances in optical technologies for data centers: A review," *Optica* **5**(11), 1354-1370 (2018). https://doi.org/10.1364/OPTICA.5.001354.
5. C. Kachris, K. Kanonakis and I. Tomkos, "Optical interconnection networks in data centers: Recent trends and future challenges," *IEEE Communications Magazine* **51**(9), 39-45 (2013). https://doi.org/10.1109/MCOM.2013.6588648.
6. M. Dayarathna, Y. Wen and R. Fan, "Data center energy consumption modeling: A survey," *IEEE Communications Surveys & Tutorials* **18**(1), 732-794 (2016). https://doi.org/10.1109/COMST.2015.2481183.
7. R. Sabella, "Silicon photonics for 5g and future networks," *IEEE Journal of Selected Topics in Quantum Electronics* **26**(2), 1-11 (2020).

https://doi.org/10.1109/JSTQE.2019.2948501.
8. P. Lin et al., "Enabling intelligent metasurfaces for semi-known input," *Progress In Electromagnetics Research* **178**, 83-91 (2023).
9. W. Zou et al., "Dispersion compensation for spoof plasmonic circuits," *Progress In Electromagnetics Research* **179**, 95-100 (2024).
10. K. Okamoto, "Wavelength-division-multiplexing devices in thin soi: Advances and prospects," *IEEE Journal of Selected Topics in Quantum Electronics* **20**(4), 248-257 (2014). https://doi.org/10.1109/JSTQE.2013.2291623.
11. C. Sun et al., "Single-chip microprocessor that communicates directly using light," *Nature* **528**(7583), 534-538 (2015). https://doi.org/10.1038/nature16454.
12. Y. Liu et al., "Arbitrarily routed mode-division multiplexed photonic circuits for dense integration," *Nature Communications* **10**(1), 3263 (2019). https://doi.org/10.1038/s41467-019-11196-8.
13. R. G. H. van Uden et al., "Ultra-high-density spatial division multiplexing with a few-mode multicore fibre," *Nature Photonics* **8**(11), 865-870 (2014). https://doi.org/10.1038/nphoton.2014.243.
14. L.-W. Luo et al., "Wdm-compatible mode-division multiplexing on a silicon chip," *Nature Communications* **5**(1), 3069 (2014). https://doi.org/10.1038/ncomms4069.
15. V. Petrov et al., "Wavefront hopping: An enabler for reliable and secure near field terahertz communications in 6g and beyond," *IEEE Wireless Communications* **31**(1), 48-55 (2024). https://doi.org/10.1109/MWC.003.2300310.
16. S. Wang et al., "26.8-m thz wireless transmission of probabilistic shaping 16-qam-ofdm signals," *APL Photonics* **5**(5), 056105 (2020). https://doi.org/10.1063/5.0003998.
17. S. R. Moon et al., "Cost-effective photonics-based thz wireless transmission using pam-n signals in the 0.3 thz band," *Journal of Lightwave Technology* **39**(2), 357-362 (2021). https://doi.org/10.1109/JLT.2020.3032613.
18. H. Zhang et al., "Aggregated 1.059 tbit/s photonic-wireless transmission at 350 ghz over 10 meters," in OSA Technical Digest P. T. H. Alexander Wai and C. Yu, Eds., *26th Optoelectronics and Communications Conference* T5A.3 (2021), https://doi.org/10.1364/OECC.2021.T5A.3.
19. S. Nellen et al., "Coherent wireless link at 300 ghz with 160 gbit/s enabled by a photonic transmitter," *Journal of Lightwave Technology* **40**(13), 4178-4185 (2022). https://doi.org/10.1109/JLT.2022.3160096.
20. S. Mao et al., "Compact hybrid five-mode multiplexer based on asymmetric directional couplers with constant bus waveguide width," *Opt. Lett.* **48**(10), 2607-2610 (2023). https://doi.org/10.1364/OL.486080.
21. W. Jiang, L. Xie and L. Zhang, "Design and experimental demonstration of a silicon five-mode (de)multiplexer based on multi-phase matching condition," *Opt. Express* **31**(20), 33343-33354 (2023). https://doi.org/10.1364/OE.502062.
22. S. Molesky et al., "Inverse design in nanophotonics," *Nature Photonics* **12**(11), 659-670 (2018). https://doi.org/10.1038/s41566-018-0246-9.
23. S. D. Campbell et al., "Review of numerical optimization techniques for meta-device design [invited]," *Opt. Mater. Express* **9**(4), 1842-1863 (2019). https://doi.org/10.1364/OME.9.001842.

24. M. M. R. Elsawy et al., "Numerical optimization methods for metasurfaces," *Laser & Photonics Reviews* **14**(10), 1900445 (2020). https://doi.org/https://doi.org/10.1002/lpor.201900445.
25. K. Yao, R. Unni and Y. Zheng, "Intelligent nanophotonics: Merging photonics and artificial intelligence at the nanoscale," *Nanophotonics* **8**(3), 339-366 (2019). https://doi.org/doi:10.1515/nanoph-2018-0183.
26. J. Jiang, M. Chen and J. A. Fan, "Deep neural networks for the evaluation and design of photonic devices," *Nature Reviews Materials* **6**(8), 679-700 (2021). https://doi.org/10.1038/s41578-020-00260-1.
27. O. Khatib et al., "Deep learning the electromagnetic properties of metamaterials—a comprehensive review," *Advanced Functional Materials* **31**(31), 2101748 (2021). https://doi.org/https://doi.org/10.1002/adfm.202101748.
28. S. So et al., "Deep learning enabled inverse design in nanophotonics," *Nanophotonics* **9**(5), 1041-1057 (2020). https://doi.org/doi:10.1515/nanoph-2019-0474.
29. A. Y. Piggott et al., "Inverse design and demonstration of a compact and broadband on-chip wavelength demultiplexer," *Nature Photonics* **9**(6), 374-377 (2015). https://doi.org/10.1038/nphoton.2015.69.
30. W. Jiang, S. Mao and J. Hu, "Inverse-designed counter-tapered coupler based broadband and compact silicon mode multiplexer/demultiplexer," *Opt. Express* **31**(20), 33253-33263 (2023). https://doi.org/10.1364/OE.500468.
31. Y. Tong, W. Zhou and H. K. Tsang, "Efficient perfectly vertical grating coupler for multi-core fibers fabricated with 193 nm duv lithography," *Opt. Lett.* **43**(23), 5709-5712 (2018). https://doi.org/10.1364/OL.43.005709.
32. A. M. Hammond et al., "Multi-layer inverse design of vertical grating couplers for high-density, commercial foundry interconnects," *Opt. Express* **30**(17), 31058-31072 (2022). https://doi.org/10.1364/OE.466015.
33. C. Q. Qingze Tan, Hongsheng Chen, "Inverse-designed metamaterials for on-chip combinational optical logic circuit," *Progress In Electromagnetics Research* **176**, 55-65 (2023).
34. V. Nikkhah et al., "Inverse-designed low-index-contrast structures on a silicon photonics platform for vector–matrix multiplication," *Nature Photonics* **18**(5), 501-508 (2024). https://doi.org/10.1038/s41566-024-01394-2.
35. K. Y. Yang et al., "Inverse-designed non-reciprocal pulse router for chip-based lidar," *Nature Photonics* **14**(6), 369-374 (2020). https://doi.org/10.1038/s41566-020-0606-0.
36. Z. Li et al., "Inverse design enables large-scale high-performance meta-optics reshaping virtual reality," *Nature Communications* **13**(1), 2409 (2022). https://doi.org/10.1038/s41467-022-29973-3.
37. K. Katanforoosh and D. J. D. a. Kunin, "Initializing neural networks," (2019).
38. S. Mao et al., "Multi-task topology optimization of photonic devices in low-dimensional fourier domain via deep learning," *Nanophotonics* **12**(5), 1007-1018 (2023).
39. M. Mitchell, *An introduction to genetic algorithms*, MIT press (1998).
40. C. M. Lalau-Keraly et al., "Adjoint shape optimization applied to electromagnetic design," *Opt. Express* **21**(18), 21693-21701 (2013). https://doi.org/10.1364/OE.21.021693.

41. K. Tsuruda, M. Fujita and T. Nagatsuma, "Extremely low-loss terahertz waveguide based on silicon photonic-crystal slab," *Opt. Express* **23**(25), 31977-31990 (2015). https://doi.org/10.1364/OE.23.031977.
42. "Method of anisotropic etching of silicon," Robert Bosch GmbH, United States (2003).